\documentclass{aastex63}

\usepackage{graphicx}
\usepackage{comment}
\usepackage{bibentry}

\shorttitle{Filaments \& IMF Bz II}
\shortauthors{Mundra et al.}
\graphicspath{{./}{figures/}}

\begin{document}
\title{Analysis of the Coronal Mass Ejections through Axial Field Direction of Solar Filaments and IMF Bz
}
\correspondingauthor{Kashvi Mundra}
\email{kashvi.mundra@gmail.com}

\author{Kashvi Mundra}
\affiliation{Lambert High School}

\author[0000-0002-0824-3109]{V. Aparna}
\affiliation{Georgia State University}

\author{Petrus Martens}
\affiliation{Georgia State University}

\begin{abstract}
In the past, there have been many studies claiming that the effects of geomagnetic storms strongly depends on the orientation of the magnetic-cloud part of the Coronal Mass Ejections (CMEs). \citet{2020ApJ...897...68A}, using Halo-CME data from 2007-2017, have shown that the magnetic field orientation of filaments at the location where CMEs originate can be effectively used for predicting the onset of geo-magnetic storms. The purpose of this study is to extend their survey by analyzing the halo-CME data for 1996-2006. The correlation of filament axial direction and their corresponding Bz signatures are used to form a more extensive reasoning for the claims presented by Aparna \& Martens before. This study utilizes SOHO EIT 195 \AA, MDI magnetogram images, KSO and BBSO H$\alpha$ images for the time period, along with ACE data for inter-planetary magnetic field signatures. Correlating all these, we have found that the trend in Aparna \& Martens' study of a high likelihood of the correlation between the axial field direction and Bz orientation, persists for the data between 1996-2006 as well.



\end{abstract}

\keywords{Space weather, CMEs, Filaments}


\section{Introduction} \label{sec:intro}

The correlation between the axial field direction of filaments and the Interplanetary Magnetic Field (IMF) Bz orientation of their corresponding Coronal Mass Ejections (CMEs) is of great importance to the astronomical community. Analyzing these quantities can give substantial clues about predicting the devastating geomagnetic storms which ravage power supplies on Earth and satellites in space. \citet{2020ApJ...897...68A} analysed filaments in CME source locations and their IMF counterparts at Earth using the dataset of halo-CMEs from 2007-2017. With the largest data-set ever used for such a study, they observationally established that a trend prevails in the relation between the direction of axial field of a CME and that of the IMF near the Earth. Current work includes the data-set taken from the years 1996-2006 and will complement the results from Aparna \& Martens, and provide a stronger base of reasoning for their results.

Similar studies have been conducted on this subject, such as \citet{2015SoPh..290.1371M}, who compared the axial orientation (without direction) of filaments on the Sun to the CMEs erupting from the locations of these filaments using flux rope fitting methods. \citet{2018SpWea..16..442P}'s study focuses upon analyzing various features of the source region at the surface of the Sun to find the axial direction and using flux rope fittings at the L1 Lagrangian point between Sun and the Earth, and uses a subset of events from 2010-2015. Their analysis has given a match rate of 55\% between the orientations at the sun and at Earth when considering a span of 45$^\circ$. \citet{1998AnGeo..16....1B} and \citet{2001ApJ...563..381Y} narrow the span even more, still suggesting that there is a high match rate between events at the Sun and at Earth. In Bothmer \& Schwenn's study, 8 out of 9 cases had the same sign of helicities at the Sun and at Earth, giving a rate of about 89\%. Yurchyshyn has given an agreement rate of 77\% of cases. \citet{2020ApJ...897...68A}, hereafter AM20, considers all the halo-CMEs between 2007 and 2017 and analyze their corresponding filaments on the Sun to get the axial field direction to compare with the Bz component of the IMF near the Earth. Their data-set is by far the largest and they classify $\sim$85\% of the cases to have a match. This has been a strong result and has the potential to provide effective space-weather predictive capabilities by only observing the filaments. The current work explores data over a different solar cycle to understand if the result is consistent over time. 
We discuss the data and the methods used in this work in the next section, analysis in section \ref{sec:analysis} and conclusions in section \ref{sec:disc}. 

\section{Data and Methods} \label{dataMethods}
\begin{figure}\label{Bz}
    \centering
    \includegraphics[trim=0 0 0 0,width=20cm,height=15cm]{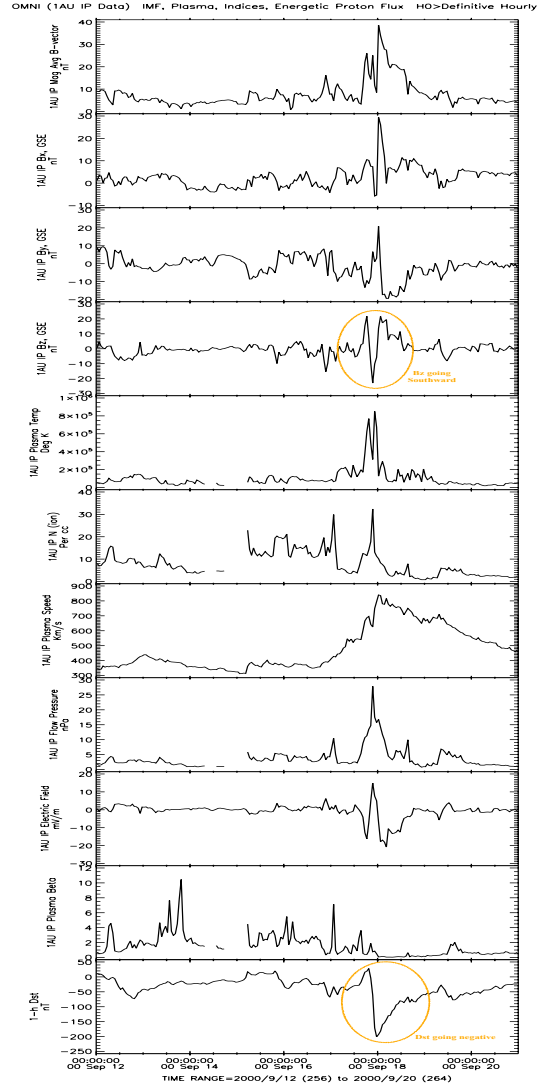}
    \caption{Plots showing the variation of the different parameters from ACE (via CDAWeb) used for analysing the IMF are shown. The data is presented in a 1 hour time format. This graph in particular is focused on the CME which occurred on September 16th, 2000. The Bz data when the CME reaches Earth shows a south-dominant orientation, so the Bz data for this date has been classified as “south”.}
\end{figure}

The methods used in AM20 is followed to conduct the analysis for this work. A brief account on the same is given below. Halo-CMEs from the 1996-2006 timeframe, listed in the Coordinated Data Analysis Workshop (CDAW) 
website\footnote{\url{https://cdaw.gsfc.nasa.gov/CME_list/halo/}} are used. 
The CMEs are analyzed to determine a correlation between the axial field direction of filaments and the IMF Bz orientation after the number of days the CME is due to reach Earth. The IMF Bz data is taken from the Advanced Composition Explorer (ACE) satellite, whose data is consolidated on the NASA CDAWeb site, along with other data such as temperature, flow speed, electric field, plasma-beta and disturbance storm-time (Dst) index. The Dst index, in particular, evaluates the disturbance in the Earth’s magnetic field giving information about the geo-effectiveness of a CME. 
The number of days for the CME to reach Earth is calculated by taking the distance between Earth and the Sun and dividing it by the space speed of an incoming CME. An example of the IMF data is shown in Fig.\ \ref{Bz}. Gonzalez \& Tsurutani (1987) determined that the causes of large geomagnetic storms are large negative changes in the Bz, exceeding -10 nT. The graph shown in the figure is thus a potential indicator of a geomagnetic storm. The Dst data in the last panel of the figure shows a sharp drop to almost -200 nT, which indicates a severe disturbance in the Earth’s atmosphere. The Bz and Dst changes occur almost at the same time, thus indicating a causal link, mentioned and predicted by \citet{1975JGR....80.4204B}. Later, \citet{2005GeoRL..3218103G} conducted a quantitative analysis of the years 1997-2002, showing the relationship between the peaks of the Bz and DST data. With reference to these previous studies, we can describe the Bz data as having a profound impact on the status of the Dst index.


To compare the IMF Bz orientation to that of the source region on the Sun, the axial field of the filament or the source active region is obtained using the method of chirality \citep{1998ASPC..150..419M}. In order to analyze the filaments on the Sun, H$\alpha$ images from the Kanzelh\"ohe catalog (KSO; \cite{2015SoPh..290..951P}) and the Big Bear Solar Observatory (BBSO) catalog were used for barbs, and the SOHO (Solar and Heliospheric Observatory) Extreme-ultraviolet Imaging Telescope (EIT; \citet{1995SoPh..162..291D}) 195{\AA}  and Michelson Doppler Imager (MDI; \citet{1995SoPh..162..129S}) magnetogram features were used for the arcades using the JHelioviewer \citep{2009CSE....11...38M} application. The article further explains the usages of these data.
To determine the chiralities of the filaments, two methods are employed. 

\begin{figure}\label{barbs}
\includegraphics[trim=-5cm 0 0 0,scale=0.7]{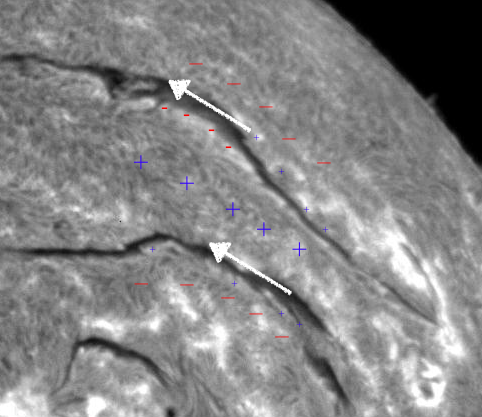}
\caption{This image was taken from the BBSO archive and depicts the north-west Sun on 02/09/2001. The  positive and negative regions obtained from the MDI Magnetogram are shown in blue and red, respectively. The pluses and the minuses in smaller font depicts the minority polarities \citep{1998ASPC..150..419M}
}
\end{figure}

1) H$\alpha$ method: using filament barbs by visually analyzing H$\alpha$ images from BBSO for the corresponding date and obtaining their direction. An example is shown in Fig.\ \ref{barbs}. The filament is a Northward-oriented specimen. This is evidenced by the direction of the barbs. The less-dominant filament regions are shown in the smaller font, and the direction of the filament axis is calculated from the positive to negative movement from these. Therefore, we can see that the positive to negative movement would render this filament to have a northward axial field. In the cases of active regions, chirality using the barbs method cannot be used. This automatically requires surveying arcade loops for that region. If no arcade loops are visible or the images are too ambiguous, no conclusion can be reached about the chirality of the filament. After the chirality is determined, the magnetic field via the MDI magnetogram is used to determine the axial orientation of the filament. This is then compared with the IMF Bz data. 

\begin{figure}\label{arcades}
\includegraphics[trim=-5cm 0 0 0,scale=0.85]{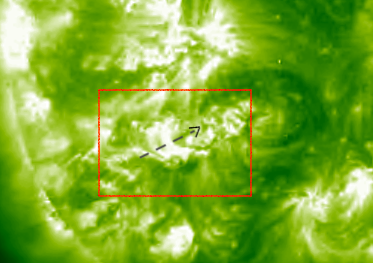}
\caption {A screenshot of the post-flare arcade loops (encased in the red box) for the flare on 2000-11-01 is shown. The black dashed line roughly represents the polarity inversion line. The corresponding magnetogram is shown in Fig.\ \ref{magnetogram}. 
}
\end{figure}

2) Arcade method: the second method is visually analyzing the post-flare arcade loops that form after a flare or a CME eruption in the source region and determining the orientation of the axis using the skew of the arcade. An example is shown in Fig.\ \ref{arcades}. Using the left or right skew of the arcade loops and their active regions \citep{1998ASPC..150..419M}, we determine the chirality of the arcade loops to be sinistral, or left-handed. Once again, following the Martin method for Axial Direction, we determine that the Axial Direction is North West, and North, taking only the vertical component for this work. The Bz data's orientation, whether it is positive dominant (north) or negative dominant (south) in the CME, is compared with this axial field direction of the filament thus obtained to determine a correlation.

The JHelioviewer application was used for analysing the arcades using SOHO EIT 195{\AA} and the MDI magnetograms for the polarities to determine the directionality of the axis. 
Similar to AM20, only HALO CMEs are considered for this study. The catalog was examined, and because this study involves heavy visual examination of the surface of the sun, any CMEs from the backside or 80-90$^\circ$ (limb) were not used for the study.

\section{Analysis} \label{sec:analysis}
\begin{figure}\label{cme_halocme_dist}
    \centering
    \includegraphics[scale=0.7]{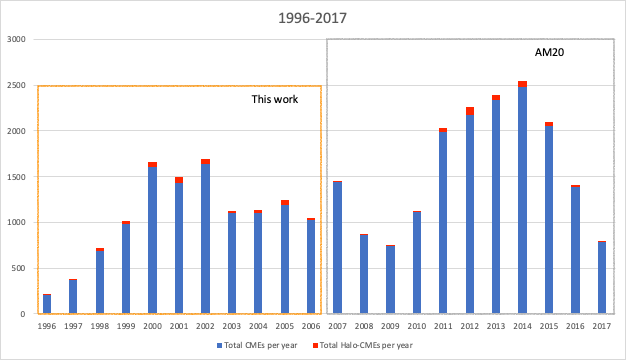}
    \caption{This chart is shown to loosely illustrate the progress of Solar Cycle 23, which peaked in 2001. The number of halo-CMEs occurring steadily increases until 2001, while the All CMEs chart shows maximums in the years 2000 and 2002. The year 2007 in the HALO CMEs chart is an outlier, as a higher-than-average amount of halo-CMEs occurred that year.}
\end{figure}

The halo-CME list for the years 1996-2006 consists of 392 halo-CMEs. Figure \ref{cme_halocme_dist} shows the yearly distribution of all the CMEs (in blue) and the halo-CMEs (in red) for both the cycles 23 and 24. The chart is shown to loosely illustrate the progress of Solar Cycles 23 and part of 24 (AM20), which peaked in 2001 and 2014, respectively. The number of halo-CMEs occurring steadily increases until 2001, while the All CMEs chart shows maximums in the years 2000 and 2002. The year 2007 in the HALO CMEs chart is an outlier, as a higher-than-average amount of HALO CMEs occurred that year after the solar maximum. The CME:halo-CME ratio was about 30:1. 

Of the 392, 84 halo-CMEs were determined for both axial field direction and the orientation of the Bz data. Of these 84, 52 CMEs had coordinating axial field direction and IMF Bz orientation, while 32 did not. Therefore, 63.09\% of the CMEs’ axial directions and IMF Bz orientations agree, while 36.91\% 's do not. The results are presented in Table 1. The CMEs that have a mismatch between the filament axial field and the the IMF Bz are thought to be because of rotation in the inter-planetary space (Yurchyshyn et al.\ 2007, \citet{2017LRSP...14....5K}, AM20). 

\begin{figure}\label{halocme_dist}
    \centering
    \includegraphics[scale=0.7]{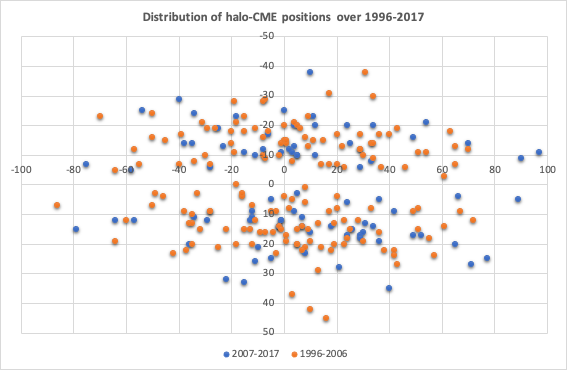}
    \caption{The graph above depicts the distribution of the CME events that were analyzed in this work and those in AM20. As seen in the graph, the vast majority of events fall between 30N and 30S, and 40E and 40W. }
\end{figure}

The remaining 308 events could not be determined because of various reasons - 54 events had a determined  axial field direction but the Bz was not determined because of a multitude of reasons, which included there being no signature in the Bz or a signature being too late or early outside of a reasonable arrival time for the CMEs. On the other hand, for 44 events, the Bz orientation was determined, however, the filament axis direction was not determined due to reasons such as the images for particular days being absent from the catalog, low resolution, filaments near the limb or the altogether absence of the filament in both the arcade loops from EIT and the H$\alpha$ images. The CMEs that had no IMF Bz signatures but were visible in terms of axial field directions may have missed the detection near Earth for many reasons, one of which could be because of the deflection of the magnetic cloud away from Earth. For example, a CME erupting from the far west can miss Earth entirely or be deflected \citep{2004SoPh..222..329W}. 47 events were completely undetermined in both the axial direction and the Bz, and 93 events were too far out to the east or the west, approaching the 90 degrees longitude, to be confidently determined as dextral or sinistral chirality. Table \ref{category} summarizes these numbers accordingly. The total compilation of the data is available on the Harvard Dataverse (link). Figure \ref{halocme_dist} shows the distribution of the determined events on the Sun for this study as well as that for AM20. This figure indicates an almost symmetric distribution of the source locations of the halo-CMEs, opposite to that found by \citet{2004SoPh..222..329W}. Further investigation into this might be necessary but is not in the scope of this paper.

The reason for the lower than average percentage of agreeing CMEs in comparison to AM20 is attributed to the solar maximum. 
The maximum of the 23rd solar cycle occurred in Nov 2001, and the cycle finished around December 2008. The years following 2001 up to 2005 are shown to be quite populated with solar flares and CMEs, which would contribute to there being less matches between the Bz and filament axis because of higher density of active regions and hence higher ambiguity in determining the chiralities (AM20).


\section{Discussion \& Conclusions} \label{sec:disc}

Through a thorough investigation, this study provides more evidence for the correlation of axial field directions of solar filaments and the Bz orientation in the corresponding CMEs. This trend has held for the years 1996-2006, and with the addition of Aparna \& Martens's previous set of 2007-2017 analysis, a total of 20 years show this correlation. 
With a set of 84 total CMEs determined in this analysis, a 63.09\% match rate between the correlations of the filament axes and the Bz orientations was found. The criteria for this research were different from previous projects such as those of \citet{2018SpWea..16..442P} and Marubashi et al.\ (2015), in that only north or south components of the field directions in the filament axis and positive or negative orientations in the IMF Bz was considered in order to simplify the process and results obtained from the project. 

Some assumptions this project has made include assuming that the speed of a CME arriving at the Earth is linear, as this is not entirely correct in every case. Working with low-resolution data  could also lead to some shortcomings. In this project, the utmost care has been taken to judge each CME with an unbiased approach, however the quality of the images can limit inspection accuracy to an extent.
Therefore, it is important to note that some ambiguous cases, upon closer inspection, were deemed to be determined. In some cases, this could also be a potential source of error, because of low resolution images or the Bz fluctuations could impair the inspection.

Previous studies such as Bothmer \& Schwenn (1998) have utilized methods in which the exact angle of the axial field directions were determined. However, with this study and Aparna \& Martens’ earlier study, we can conclude that just analyzing the north and south orientation of filaments and Bz signatures is sufficient to provide sound data to prepare for a geomagnetic storm. To favor the north-south approach further, we note that in the case of severe geomagnetic storms in the past, such as the Halloween Storms of 2003 (Reference), many of the CMEs that were found to correlate to the disruptions in the Bz and DST (which were in the negative range) arose out of filaments with a southward axial field direction.

\begin{table}[]\label{distribution}
    \centering
    \begin{tabular}{c|c|c}
         IMF Bz & North & South  \\
         North & 32 & 19 \\
         South & 13 & 20 \\
         & & \\
         Total & 84 & \\
         Agree & 53\% & 63.09\% \\
         Disagree & 31\% & 36.91\%  \\
    \end{tabular}
    \caption{Filament chirality and IMF Bz distribution.}
\end{table}

\begin{figure}\label{magnetogram}
    \centering
    \includegraphics[scale=0.8]{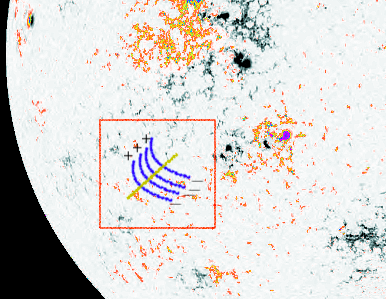}
    \caption{This image is the MDI Magnetogram rendering of 2000-11-01 Sun. Orange represents negative and grey positive polarities. The active region shown in Fig.\ \ref{arcades} has been encased in the red box. The EIT 195{\AA} arcade loops have been drawn in blue over the polarity inversion line (yellow-brown).
}
\end{figure}

\begin{table}[]\label{category}
    \centering
    \begin{tabular}{c|c}
       Determined:  & 84 \\
       Cannot be determined due to absence of filament axis (but Bz is there):  & 54 \\
       Cannot be determined due absence of Bz (but filament axis is there): & 44 \\
       That are backsides: & 71 \\
       That are not determined in any way: & 47 \\
       Limb 90: & 93 \\
       Total & 393 
    \end{tabular}
    \caption{The table categorizes the data based on determined and non-determined events.}
\end{table}


\pagebreak

\section{Acknowledgements}

We would like to thank CDAW team for providing the CME and Halo-CME list on their website, without which this research would not have been possible. We thank OMNIWeb for their effort in creating a common database for such parameters from various spacecrafts.
We thank the ACE instrument team and the ACE Science Center for providing the ACE data.  We also extend our thanks to the SOHO/MDI and the EIT consortia for the data. SOHO is a project of international cooperation between ESA and NASA.

Kanzelh\"ohe Observatory (KSO), from University of Graz, Austria and the Big Bear Solar Observatory (BBSO) provided the H$\alpha$ data. BBSO operation is supported by NJIT and US NSF AGS-1821294 grant. GST operation is partly supported by the Korea Astronomy and Space Science Institute, the Seoul National University, and the Key Laboratory of Solar Activities of Chinese Academy of Sciences (CAS) and the Operation, Maintenance and Upgrading Fund of CAS for Astronomical Telescopes and Facility Instruments. We are also grateful to J.H.\ King and N.\ Papatashvilli of Adnet Systems, NASA GSFC and CDAWeb for providing the interplanetary plasma data.

\bibliography{sample63}{}
\begin{itemize}
\bibitem[Aparna \& Martens(2020)]{2020ApJ...897...68A} Aparna, V. \& Martens, P.~C.\ 2020, \apj, 897, 68. 
\bibitem[Bothmer \& Schwenn(1998)]{1998AnGeo..16....1B} Bothmer, V. \& Schwenn, R.\ 1998, Annales Geophysicae, 16, 1. doi:10.1007/s00585-997-0001-x
\bibitem[Burton et al.(1975)]{1975JGR....80.4204B} Burton, R.~K., McPherron, R.~L., \& Russell, C.~T.\ 1975, \jgr, 80, 4204. 
\bibitem[Delaboudini{\`e}re et al.(1995)]{1995SoPh..162..291D} Delaboudini{\`e}re, J.-P., Artzner, G.~E., Brunaud, J., et al.\ 1995, \solphys, 162, 291
\bibitem[Gonzalez \& Echer(2005)]{2005GeoRL..3218103G} Gonzalez, W.~D. \& Echer, E.\ 2005, \grl, 32, L18103. 
\bibitem[Gonzalez & Tsurutani(1987)]{1987P&SS...35.1101G} Gonzalez, W.~D. \& Tsurutani, B.~T.\ 1987, \planss, 35, 1101. 
\bibitem[Kilpua et al.(2017)]{2017LRSP...14....5K} Kilpua, E., Koskinen, H.~E.~J., \& Pulkkinen, T.~I.\ 2017, Living Reviews in Solar Physics, 14, 5 
\bibitem[Martin (1998)]{1998ASPC..150..419M} Martin, S.~F.\ 1998, IAU Colloq.~167: New Perspectives on Solar Prominences, 150, 419 
\bibitem[Marubashi et al.(2015)]{2015SoPh..290.1371M} Marubashi, K., Akiyama, S., Yashiro, S., et al.\ 2015, \solphys, 290, 1371. 
\bibitem[M\"uller et al.(2009)]{2009CSE....11...38M} M\"uller, D., Fleck, B., Dimitoglou, G., et al.\ 2009, Computing in Science and Engineering, 11, 38
\bibitem[Palmerio et al.(2018)]{2018SpWea..16..442P} Palmerio, E., Kilpua, E.~K.~J., M{\"o}stl, C., et al.\ 2018, Space Weather, 16, 442. 
\bibitem[P{\"o}tzi et al.(2015)]{2015SoPh..290..951P} P{\"o}tzi, W., Veronig, A.~M., Riegler, G., et al.\ 2015, \solphys, 290, 951
\bibitem[Scherrer et al.(1995)]{1995SoPh..162..129S} Scherrer, P.~H., Bogart, R.~S., Bush, R.~I., et al.\ 1995, \solphys, 162, 129
\bibitem[Wang et al.(2004)]{2004SoPh..222..329W} Wang, Y., Shen, C., Wang, S., et al.\ 2004, \solphys, 222, 329. doi:10.1023/B:SOLA.0000043576.21942.aa
\bibitem[Yurchyshyn et al.(2001)]{2001ApJ...563..381Y} Yurchyshyn, V.~B., Wang, H., Goode, P.~R., et al.\ 2001, \apj, 563, 381.
\bibitem[Yurchyshyn et al.(2007)]{2007AdSpR..40.1821Y} Yurchyshyn, V., Hu, Q., Lepping, R.~P., et al.\ 2007, Advances in Space Research, 40, 1821.
\end{itemize}
\bibliographystyle{aasjournal}


\end{document}